\documentclass{aa}  

\usepackage[utf8]{inputenc}
\usepackage{keyval}
\usepackage{graphicx}
\usepackage{txfonts}
\usepackage[colorlinks=true,linkcolor=red]{hyperref}
\usepackage{natbib}
\DeclareUnicodeCharacter{200B}{}
\begin{document}

   \title{LRPayne: Stellar parameters and abundances from low-resolution spectra }

   \authorrunning{Nagaraj Vernekar et al.}
 \titlerunning{ LRPayne: Stellar parameters and abundances from low-resolution spectra }

   \author{Nagaraj Vernekar
   \inst{1,2}
    \and
   Lorenzo Spina \inst{2}
   \and
   Sara Lucatello\inst{2}
   \and
   Carmelo Arcidiacono\inst{2}
   \and
   Luca Cortese\inst{1,2}
   \and
   Matteo Simioni\inst{2}
   \and
   Andrea Balestra\inst{2}
   }

   \institute{
   Dipartimento di Fisica e Astronomia, Universit\'a di Padova, vicolo dell'Osservatorio 2, 35122 Padova, Italy \\
  \email{nagarajvernekar30@gmail.com, nagaraj.vernekar@inaf.it}
   \and
  INAF - Ossevatorio Astronomico di Padova, vicolo dell'Osservatorio 5, 35122 Padova, Italy}

   \date{Received XXXX; accepted XXXX}

% \abstract{}{}{}{}{} 
% 5 {} token are mandatory
 
  \abstract
   {}
   {This paper introduces LRPayne, a novel algorithm designed for the efficient determination of stellar parameters and chemical abundances from low-resolution optical spectra, with a primary focus on data from large-scale galactic surveys such as WEAVE.}
   {LRPayne employs a model-driven approach, utilising a fully connected artificial neural network (ANN), trained on a library of 70,000 synthetic stellar spectra generated using iSpec with 1D MARCS model atmospheres and the Turbospectrum synthesis code. The network is trained to predict normalized flux given stellar labels (T$_{\rm eff}$, log(g), [Fe/H], $v_{mic}$, $v_{max}$ and $v$\,sin\,i, and 24 individual elemental abundances). Stellar parameters are subsequently derived from observed spectra by finding the best-fit synthetic spectrum from the ANN using a $\chi^{2}$ minimisation technique. The method operates on spectra degraded to a resolution of R=5000 covering the wavelength range 4200-6900 Å}
   { Internal accuracy tests on synthetic spectra show a median interpolation error of less than 0.13 $\%$ for 90 $\%$ of the validation sample. The method accurately recovers most input labels from synthetic spectra, even at a signal-to-noise ratio (S/N) of 20, with some expected challenges for elements like Li, K, and N. Validation on observed spectra of 25 Gaia FGK benchmark stars and 42 metal-poor stars reveals good agreement with literature values. For stellar parameters, mean differences are 22±87 K for T$_{\rm eff}$​, 0.19±0.23 dex for log(g), and 0.01±0.17 dex for [Fe/H]. Abundances for elements like Na, Mg, Si, and most Fe-peak elements (Cr, Ni, V, Sc) are well-recovered. Challenges are noted for oxygen, manganese in metal-rich giants, aluminium in metal-poor stars and dwarfs, and for deriving log g in hot metal-poor dwarfs, partly due to non local thermodynamic equilibrium effects and line characteristics.}
   {LRPayne demonstrates the possibility of extracting precise stellar parameters and chemical abundances from large amounts of low-resolution spectra. Its good performance for different kinds of stars makes it well suited for current and future large surveys. The abundance results from LRPayne will be very useful for studying stellar nucleosynthesis and the chemical evolution of our Galaxy on a large scale.}


   \maketitle
%
%-------------------------------------------------------------------

\section{Introduction}

The era of large-scale spectroscopic surveys has transformed stellar astrophysics, thanks to their unprecedented data volumes. Current and upcoming ground-based surveys such as WEAVE\footnote{WHT Enhanced Area Velocity Explore} \citep{weave2012,weave2024}, 4MOST\footnote{4-metre Multi-Object Spectroscopic Telescope} \citep{4MOST2019}, DESI\footnote{Dark Energy Spectroscopic Instrument} \citep{desi2016}, and the completed APOGEE\footnote{Apache Point Observatory Galactic Evolution Experiment} \citep{apogee2017}, Gaia-ESO \citep{gaiaeso2012}, RAVE\footnote{The Radial Velocity Experiment}\citep{rave2006}, LAMOST\footnote{Large Sky Area Multi-Object Fiber Spectroscopic Telescope} \citep{2012RAA....12.1197C} and GALAH\footnote{GALactic Archaeology with HERMES} \citep{galah2015} surveys are collectively targeting 10$^6$-10$^7$ stars, providing an unparalleled view of Galactic structure, stellar populations, and chemical evolution. For instance, WEAVE alone is expected to observe roughly 5 million objects over its operational lifetime, while 4MOST aims to obtain spectra for more than 10 million objects across multiple Galactic components. Special mention should be given to the Gaia and Euclid missions, which have/will collected millions of spectra—mostly at extremely low resolutions (Gaia BP/RP; \citealt{2023A&A...674A...3M} and Euclid NISP; \citealt{2025A&A...697A...3E})—along with a few million spectra at medium resolution (Gaia RVS; \citealt{2023A&A...674A...1G}).

Though traditional spectroscopic analysis methods are generally accurate, they are not scalable to handle such enormous datasets within reasonable time-frame and computation power. Therefore, machine learning techniques have emerged as the natural solution to these challenges. Recent years have witnessed a surge in the development and application automated techniques to stellar spectroscopy.  Notable examples include The Cannon \citep{melissa2015}, which popularised the use of data-driven models for large-scale spectroscopic analysis, StarNet \citep{starnet}, which employed convolutional neural networks for stellar parameter determination, and The Payne \citep{Payne}, which introduced a model-driven approach using synthetic spectra to train neural networks for full spectral fitting. 

Although these efforts have demonstrated the huge potential of machine learning in stellar spectroscopy, most have focused primarily on high-resolution spectroscopic data ($R\,\ge\,20,000$). However, the landscape of upcoming surveys reveals a different reality: the vast majority of spectroscopic data will be obtained at low to moderate resolution ($R\,\sim\,2,000-10,000$). This shift toward lower resolution is driven by practical considerations: higher spectral multiplexing capabilities, increased survey efficiency, and the ability to observe fainter targets and hence probe a larger volume within reasonable integration times. Surveys like WEAVE's Low Resolution (LR) mode ($R\,\sim\,5,000$), 4MOST's low-resolution spectroscopy ($R\,\sim\,4000 - 7700$), and similar modes in other facilities will generate the bulk of future spectroscopic data. 

A major scientific goal for development of automated methods is to extract detailed chemical information for the vast number of stars observed by these surveys. Reliable abundance estimation from low-resolution spectra will make it possible to investigate stellar nucleosynthesis and trace the chemical enrichment of the Milky Way across different Galactic components, shedding light on the processes involved in their formation. 

In this paper, we present LRPayne\footnote{\url{https://github.com/nagarajvernekar30/lrPayne}}, a machine learning algorithm specifically designed for the efficient determination of stellar parameters and chemical abundances from low-resolution optical spectra. Our approach builds upon the model-driven philosophy of The Payne but incorporates several tweaks tailored to the low-resolution regime. The paper is structured as follows: Section 2 describes our methodology, including the observed data used for validation, the synthetic spectral library, and the neural network architecture. Section 3 presents our results, including internal accuracy tests on synthetic spectra and validation using benchmark stars and metal-poor stars with well-established parameters from the literature. We conclude with a summary of our main findings and discuss the implications for large-scale spectroscopic surveys.

\section{Method}
\subsection{Architecture}

For the algorithm, we adopted a model-driven approach of The Payne\footnote{\url{https://github.com/tingyuansen/The_Payne}} but with few modifications. The Payne is a machine learning-based approach to full spectral fitting used for determination of stellar labels. At its core, Payne uses fully connected Artificial Neural Network (ANN) as an spectral interpolator to estimate the flux values at each wavelength point given a set of stellar parameters. This is a perfect solution for analysis of low-resolution spectra due to its ability to simultaneously fit multi-dimensional labels by making use of the whole spectrum. Thus allowing direct estimation of the uncertainties on the derived parameters as well as characterisation of the covariances between different labels arising due to line blending.

In \cite{Payne}, they employ a fully connected four-layer ANN consisting of an input layer, two hidden layers, and an output layer. The number of neurons in the input and output layers is dictated by the number of parameters used for the analysis (hereafter referred to as labels) and the number of wavelength points in the final output model (hereafter referred to as pixels). However, the number of neurons in the hidden layers and the number of layers itself can vary depending on the complexity of the dataset. In \cite{Payne}, they determined that two hidden layers with 300 neurons each was a good balance ensuring reliable interpolation without the risk of overfitting. 

\begin{figure*}[h]
    \centering
    \includegraphics[width=0.85\linewidth]{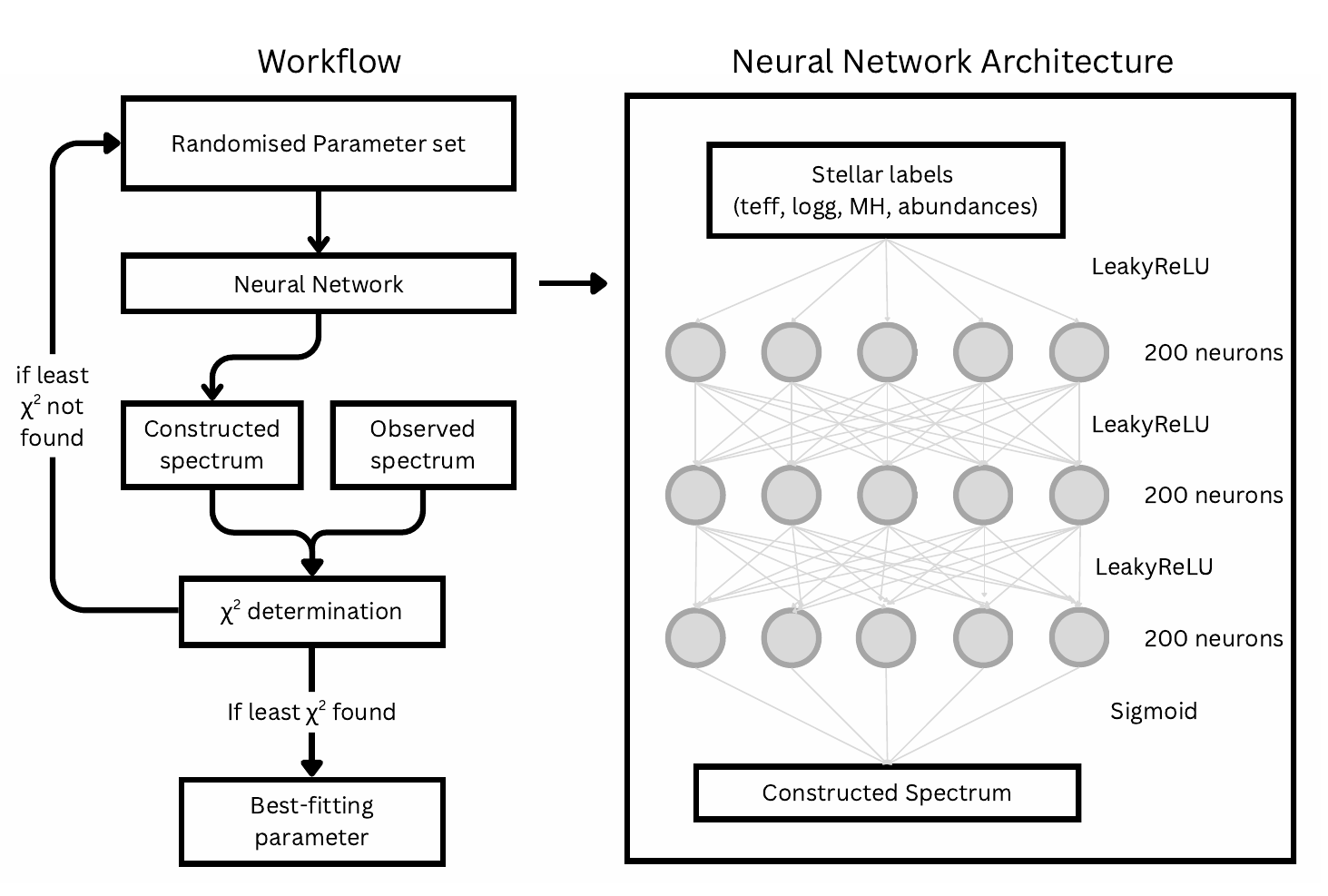}
    \caption{Pictorial representation of the workflow of LRPayne (\textit{left}) and architecture of the artificial neural network used with LRPayne (\textit{right}). }
    \label{fig:NNarchitecture}
\end{figure*}

In our case, after extensive testing, we settled with a five-layer model where each of the three hidden layers has 200 neurons. Between the first four layers, we use a LeakyReLU activation function ($LeakyReLU(x) = max(0,x)\,+\,slope\,*\,min(0,x)$) and a sigmoid activation function ($s(x) = (1\,+\,e^{-x})^{-1}$) between the last hidden layer and output layer. LeakyReLU function helps in the prevention of vanishing gradient problem where the values of the weights get quite small, leading to extremely slow convergence or stagnation of the training. The sigmoid function ensures the output values are always between 0 and 1, which is necessary for a normalised spectrum. The complete workflow and architecture of the neural network are illustrated in Figure \ref{fig:NNarchitecture}, with the model implemented in {\em TensorFlow} \citep{tensorflow2015-whitepaper}.

\subsection{Training}
For the training of the neural network, we use a sample of 70\,000 synthetic spectra which are divided into training and validation set with a 80-20\% split. The model uses stochastic optimizer ADAM \citep{adam2014} with an initial learning rate of 0.001 and a learning rate scheduler that reduces the rate by a factor of 0.5 for every 10 steps of unchanged loss. For the loss, we use the mean absolute percentage error to help with the gradient descent. The training is performed in two steps within each iteration, i.e. forward and backward propagation. Before the start of the first forward propagation, we assign random weights to each layers. This is done using layer initializers, which are set of instructions that define how these random weights are set (in our case, we use he-initialiser \citep{2015arXiv150201852H} for all three hidden layers). Up on initialisation, the ANN uses the input (taken from training set) and randomised weights to calculate a prediction that is compared with the desired output to obtain the loss. Then in backward propagation step, this loss is used to calculate the gradient with respect to each weight and bias, which is then updated to a new value depending on the type of optimizer and learning rate used. The new weights are then used for the next forward propagation and so on. The ANN uses validation sample to keep a track of the gradient descent in order to avoid over-fitting along with serving as a generalization test. As the weights of the ANN are optimised using training data, testing the optimisation on unseen validation sample provides the user with an unbiased estimate of the model's performance on unseen generalised data.

Once we have a trained model, we can input a random set of parameters into the model to obtain a normalised spectrum which is referred to as the constructed spectrum. We use the curve$\_$fit routine within the Scipy package of python to perform a simple $\chi^{2}$ minimisation between the constructed spectrum and the observed spectrum to determine the best-fit stellar parameters. To ensure that curve$\_$fit does not fall within local minima, we adopt a tolerance of $10^{-10}$, which was found to be a good compromise between reliable convergence and computation time. 

\subsection{Masking}

\begin{figure*}[h]
    \centering
    \includegraphics[width=0.85\linewidth]{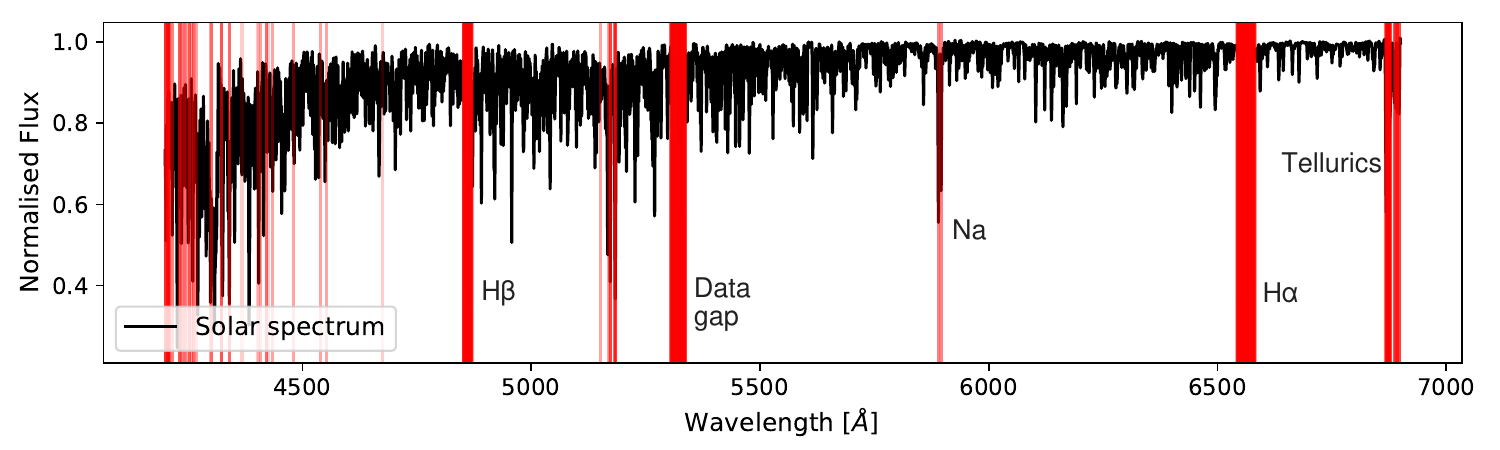}
    \caption{Solar spectrum (black) with masked pixels show with red vertical lines. Important masked lines and data gap are shown in text.  }
    \label{fig:masking}
\end{figure*}

One major problem that arises with using neural network trained on synthetic models to fit observed spectrum, is the inability of the network to reproduce any particular observed spectrum perfectly. This is likely to be due to incomplete and approximate treatment of the physics relevant to stellar atmospheres such as NLTE and 3D effects, incomplete linelists, and inaccurate atomic parameters. Therefore, we followed the method described in \cite{Payne} of masking bad pixels when fitting. To this end, we compared the high S/N solar observed spectrum and the constructed spectrum (constructed using solar values for the labels) and masked any pixel that has a deviation larger than 0.03. Although we tested using both the Sun and Arcturus spectra to identify bad pixels, the results were comparable, and hence we adopted the simpler approach of using only the solar spectrum. Along with bad pixels, we also masked the hydrogen Balmer lines as well as the gap in the data as shown in Figure \ref{fig:masking}. The Balmer lines were excluded due to considerable discrepancies between the synthetic and observed spectra in these regions, which can strongly affect the chi-square minimization, thus derailing the fitting process. It was therefore preferable to lose this information and rely on metallic lines such as Fe and Ti for parameter estimation. This lead to the masking of 540 pixels or about 5 $\%$ of the spectrum. We note that, within low-resolution, masking plays quite a crucial role as a stringent mask will remove excessive amount of information, whereas, a more lenient mask would fail to remove problematic pixels that can derail the fitting. One should make sure to find a balance between removing bad pixels and losing excessive information.

\subsection{Synthetic Spectra}

The synthetic stellar library was generated using 1D MARCS atmospheric models \citep{marcs} and Turbospectrum synthesis code \citep{1998A&A...330.1109A,turbospectrum} within iSpec. The model atmospheres are 1D plane-parallel models with LTE approximations. The linelist used for the analysis was created by supplementic the atomic lines from GES linelist with lines from VALD. 

We computed 70\,000 synthetic spectra with a total of 30 varying parameters, i.e. 6 physical parameters : T$_{\rm eff}$, log\,g, [Fe/H], $v_{mic}$, $v_{mac}$  and $v$\,sin\,i with the rest 24 parameters being atomic abundances (Li, C, N, O, Na, Al, Mg, Si, Ti, K, Ca, Mn, Cr, Ni, S, V, Sc, Ba, La, Eu, Y, Zn, Zr, Cu). The physical parameter space was selected to ensure full coverage of FGK stars in all evolutionary stages. For this reason, the bounds for the effective temperature T$_{\rm eff}$ are at 4000 and 8000 K, with log\,g between 0.0 and 5.0 dex and metallicities between -2.5 and 0.5 dex. The values for these parameters were randomly drawn within the above mentioned bounds, with T$_{\rm eff}$ and log\,g values compared to set of MIST stellar evolutionary tracks of FGK stars to include physically plausible combinations. The value of $v_{mic}$ and $v_{max}$ were calculated for each star using the values of T$_{\rm eff}$, log\,g and [Fe/H] using empirical relations (refer to \cite{ispec2014} for more information). As the analysis was carried out on low-resolution spectrum, we selected the range of $v$\,sin\,i to be between 2 and 20 km\,s$^{-1}$ with steps of 2 km\,s$^{-1}$. All the abundances [X/Fe] were randomly drawn from a range of -0.5 to 0.5 dex with solar abundances taken from \cite{Asplund2009}.  The ranges of all the parameters have been summarised in Table \ref{tab:parameter_space}. Once the spectra are created, they are degraded into desired resolution of 5000 and resampled to match the wavelength grid of WEAVE LR (four wavelength pixels per $\AA$). 

\begin{table}[]
    \centering
    \begin{tabular}{l  l}
    \hline
    
 Labels & Range  \\
 \hline
 \hline
  T$_{\rm eff}$ & [4000, 8000] (K) \\
  Log\,g  &  [0.0, 5.0] (dex) \\
 $ \rm [Fe/H]$  &  [-2.5, 0.5] (dex) \\
  $v_{mic}$ & [0, 15] (km\,s$^{-1}$) \\
  $v_{max}$ &  [0, 50] (km\,s$^{-1}$) \\
  $v$\,sin\,i & [2,20] (km\,s$^{-1}$) \\
  $ \rm [X/Fe]$ & [-0.5, 0.5] (dex) \\
  \hline
    \end{tabular}
    \caption{Parameter space used in this study}
    \label{tab:parameter_space}
\end{table}

\subsection{Observed Spectra}

 \subsubsection{Data}
The observed spectroscopic data was used for the astrophysical verification of the algorithm. As the algorithm was primarily designed to analyse low-resolution WEAVE data, it is important for the observed data to have similar properties to that of WEAVE data. The resolution could be handled easily as any spectra observed in higher-resolution can be degraded to a resolution of choice. The biggest hurdle was to find data with the full optical coverage like WEAVE (3660 to 9590 $\AA$). This vastly narrowed the instruments from which the data can be used. Therefore, instead of this full range, we conduct our analysis from 4200 till 6900 $\AA$ because it is covered by prominent instruments such as HARPS \citep{harps}, HARPS-N \citep{harps-n}, FEROS \citep{feros}, NARVAL \citep{narval}, etc and contains most of the usable key features within the optical spectra . With this availability, we divided our verification sample into two: Gaia FGK benchmark stars and metal-poor stars taken from \cite{bensby2014}.

Gaia FGK benchmark stars are a set of calibration stars defined in \cite{heiter2015} and \cite{hawkin2016} that covers a wide range of parameter space and have been analysed ensuring minimal dependency of stellar parameter estimations on spectroscopic and atmospheric models. Among the 34 stars, we include 23 stars in our analysis due to them falling within our parameter space and having data with sufficient wavelength coverage. The literature values of the stellar parameters for these stars were taken from \cite{heiter2015} and \cite{jofre2014}. The observations were carried out using instruments such as HARPS, NARVAL and ESPaDOnS \citep{espadons} with resolutions between 65\,000 and 115\,000, with S/N between 100 and 1000 per $\AA$ and were downloaded from Gaia FGK spectral library \citep{blanco2014}\footnote{https://www.blancocuaresma.com/s/benchmarkstars}. The abundances available for these stars are Mg, Si, Ca, Ti, Sc, V, Mn, Cr, Co and Ni and the values were obtained from \cite{jofre2015} and \cite{jofre2017}.

In order to increase the verification sample and have a good coverage of the entire parameter space (especially the metal-poor region), we also included stars from \cite{bensby2014} (hereafter referred to as metal-poor stars). An additional advantage of using this study is the availability of Na abundance, which is one of the key abundances for this analysis due to its importance in identification of first and second generation stars within clusters. All the spectroscopic observations for these stars were carried out using the HARPS spectrograph and are available in the ESO archive. These spectra cover a wavelength range of 4200 to 6900 $\AA$ (with a gap of 30 $\AA$ at 5300 $\AA$) and have a lower S/N compared to the benchmark stars between 50 and 300 per $\AA$. From \cite{bensby2014}, we obtained stellar parameters and abundances of O, Na, Mg, Al, Si, Ca, Ti, Cr, Fe, Ni, Zn, Y.

\subsubsection{Pre-processing}

The spectra for different stars were obtained from three different instruments at three different resolutions and wavelength coverage. In order to homogenize the entire sample to match the required specifications, we created a pipeline to perform the pre-processing. This pipeline heavily relies on functions defined within iSpec tool\footnote{iSpec is a software written in python to perform spectroscopic analysis such as radial velocity estimation, determination of stellar parameters and abundance estimation. } \citep{ispec2014,ispec2019}.  The pipeline is as follows with iSpec functions used in brackets:

\begin{itemize}
    \item Extract wavelength, flux and spectral resolution from the fits header. 
    \item Spectral range of the spectrum is narrowed to 4200 and 6900 $\AA$ with the rest being removed. 
    \item Radial velocity estimation is performed using cross-correlation method with the high S/N solar spectrum as a template. See \cite{ispec2014} for more details. Upon estimation, the spectrum is corrected to restframe ($determine\_radial\_velocity\_with\_mask$ and $correct\_velocity$).
    \item Multiple radial velocity corrected spectra for the same star observed using same instrument are merged together to enhance the S/N of the resultant spectrum. For this, we simply take the median or the mean (in case of a star has only two spectra) of the flux at each wavelength point .
    \item For continuum normalisation, the spectrum was first divided into 10 segments with each segment being about 270 $\AA$ except for one due to gap in the data. Within each segment, the continuum points are estimated using a median and maximum filter, ensuring the strong absorption lines are ignored (wavelength ranges of several strong absorption lines such as hydrogen balmer lines are provided). With the points, multiple spline functions with a degree equal to one are used to model the continuum. The flux is divided by this continuum to obtain the normalized spectrum ($fit\_continuum$, $normalize\_spectrum$). 
    \item In several stars, the estimated continuum is slightly higher than it should be. It gets worse with lower S/N spectrum. To remove this, the entire spectrum is divided by the offset, which is estimated by taking the median of the flux constituting the continuum .
    \item The co-add spectrum is still at high-resolution, therefore, the resolution of the spectrum is degraded to R = 5000 through convolution with a Gaussian ($convolve\_spectrum$).
    \item The convolved spectrum is resampled using linear interpolation with 4 points per $\AA$, thus making each spectrum about 10799 pixels in length. This is in agreement with the low-resolution spectrum of WEAVE survey. 
\end{itemize}

These convolved spectra are used as the verification sample to test the reliability of the neural network's performance on real observed spectra.

\section{Results}

\subsection{Internal Accuracy}

\begin{figure*}
    \centering
    \includegraphics[width=0.99\linewidth]{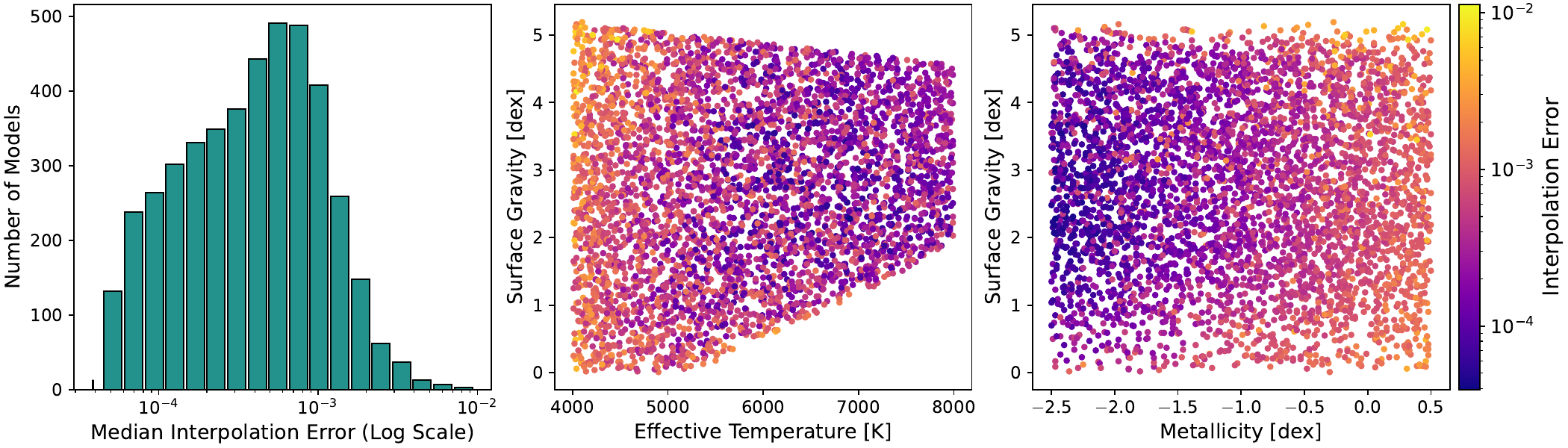}
    \caption{\textit{Left: }Histogram representing median interpolation error of the ANN for 5000 synthetic models; \textit{Middle: }Co-relation of interpolation error on effective temperature and surface gravity of the synthetic models; \textit{Right: }Co-relation of interpolation error on metallicity and surface gravity of the synthetic models.}
    \label{fig:interpolation}
\end{figure*}

In order to verify the performance of LRPayne, we used the same two tests used by \cite{Payne}.  An additional 5000 synthetic spectra were synthesised to be used for the two internal accuracy tests which will be referred to as the synthetic verification sample. These spectra were unseen by the neural network during its training. First test is to examine the interpolation accuracy of the neural network by comparing a synthetic spectrum generated using turbospectrum with the one interpolated by the neural network for the same labels. The results for the 5000 synthetic validation sample is shown in Figure \ref{fig:interpolation}, where the first panel shows the median of the difference between the constructed spectrum and the synthetic spectrum, and the next two panels show the interpolation accuracy throughout the parameter space. The neural network has an interpolation error of less than 1.3 * 10$^{-3}$ for 90 $\%$ of the sample, which translates to an error of less than 0.13 $\%$. The maximum interpolation error was found to be 9.5 *  10$^{-3}$ (less than 1 $\%$) for cool metal-rich dwarfs. This trend is clearly seen in second and third panel of Figure \ref{fig:interpolation}, where cooler metal-rich stars have relatively worse interpolation error compared to the rest. This is expected as the spectra from these stars are rich in spectral lines, often blended together at low-resolution, thus making it difficult to accurately interpolate. This result shows the interpolation accuracy of the neural network is quite high. 

\begin{figure*}
    \centering
    \includegraphics[width=0.99\linewidth]{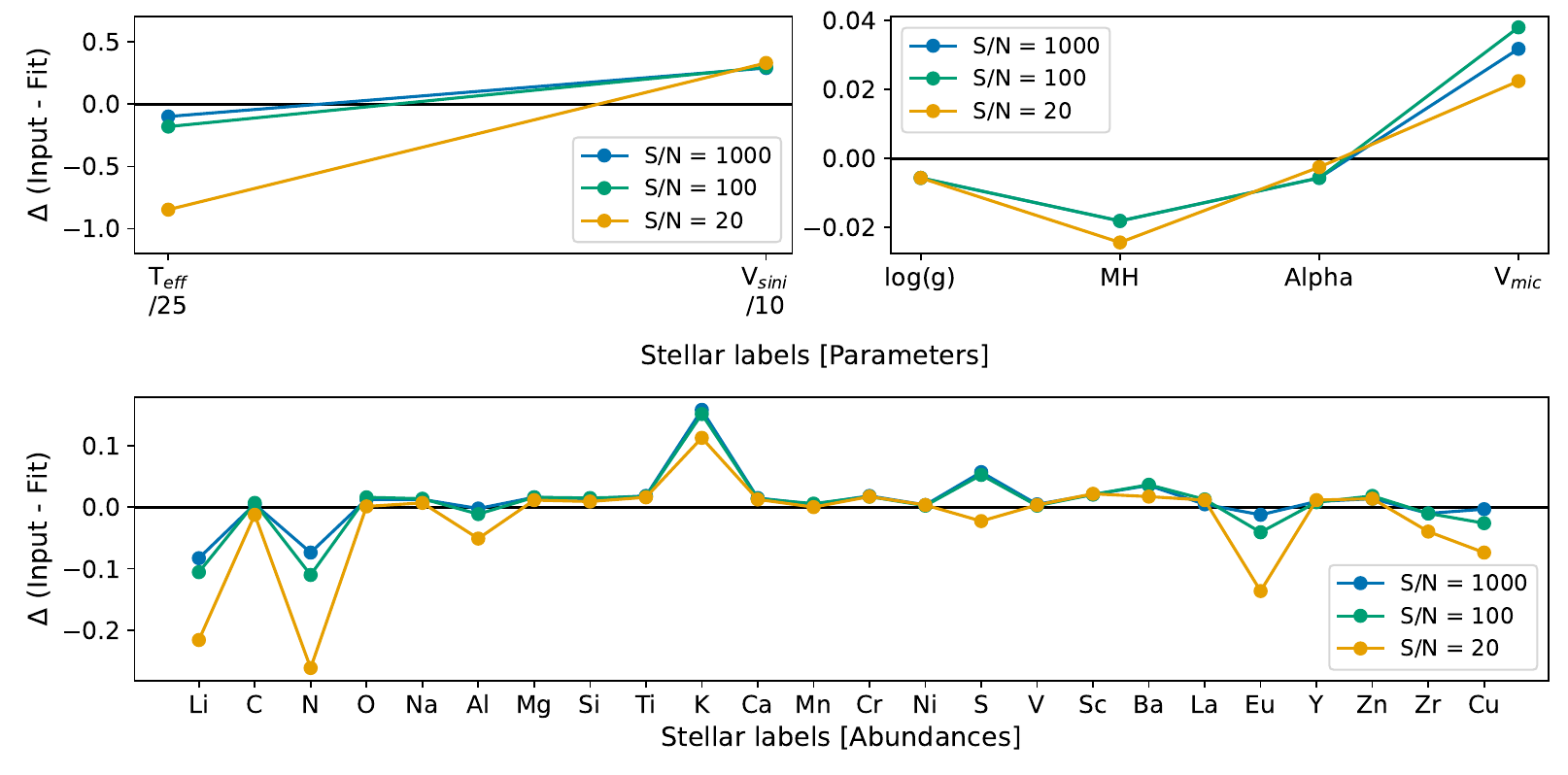}
    \caption{Comparison of input labels with respect to the values recovered by LRPayne though fitting multiple synthetic spectra at different S/N (Green : S/N = 20, Red : S/N = 100 and Black : S/N = 1000). Note: Only 14 elements [O, Na, Al, Mg, Si, Ti, Ca, Mn, Cr, Ni, V, Sc, Ba and Y] have been validated using observed data (See Section \ref{Sec:abun}). For other elements, this result represents the upper limit on the sensitivity of LRPayne}
    \label{fig:sensitivity}
\end{figure*}

The second test is to examine how well the entire workflow can recover the labels from synthetic models using $\chi^{2}$ minimisation. For this, the same synthetic validation sample is used but at three different signal-to-noise ratios (S/N), i.e. 1000, 100 and 20. We use iSpec's python function ($add\_noise$) to degrade the S/N of the synthetic spectra. Even though the added noise does not properly represent the real noise in an observed spectrum, we can use them to establish an upper limit to the accuracy of the tool to recover the labels at different S/N. The median of the difference between the input value and recovered values for 30 parameters from fitting 5000 spectra is shown in Figure \ref{fig:sensitivity}. From the results of both 1000 and 100 S/N, we see a high accuracy in most of the parameters except for V$_{\rm sini}$, Li and K. For Li and K, it is expected to not perform well due to availability of only one or two lines that are usually weak and blended, and for V$_{sini}$, we see a median discrepancy of about 2.5 km\,s$^{-1}$. But for S/N = 20, we see a relatively larger discrepancy in T$_{\rm eff}$ ($\sim$ 20 K) , Li ($\sim$ 0.22 dex), K ($\sim$ 0.1 dex), N ($\sim$ 0.28 dex), and Eu ($\sim$ 0.15 dex). Given the resolution, noisy spectra and unavailability of good lines, such a decrease in accuracy is expected.

The above tests demonstrates the capabilities of LRPayne to handle model interpolation as well as extraction of stellar labels. The dispersions obtained in these tests represents the lower limit on the uncertainties of the parameters, but are expected to increase during analysis of observed spectra due to effects such as incomplete physics in synthetic regime, inaccurate continuum normalisation and complex noise profile.

\subsection{Benchmark and metal-poor stars}
Upon satisfactory results from the internal accuracy tests, we test the ability of LRPayne to reliability recover stellar labels by fitting real observed spectra using an observed validation sample made up of 23 Gaia benchmark stars and 41 metal-poor stars. As mentioned above, spectral data for different stars come from different instruments, but they have been preprocessed to match the instrumental characteristics of the low-resolution data from the WEAVE spectrograph. The results for stellar parameters and abundances are shown in Figure \ref{fig:parameters} and \ref{fig:abundances}, respectively.

\subsubsection{Parameters}

\begin{figure}
    \centering
    \includegraphics[width=1\linewidth]{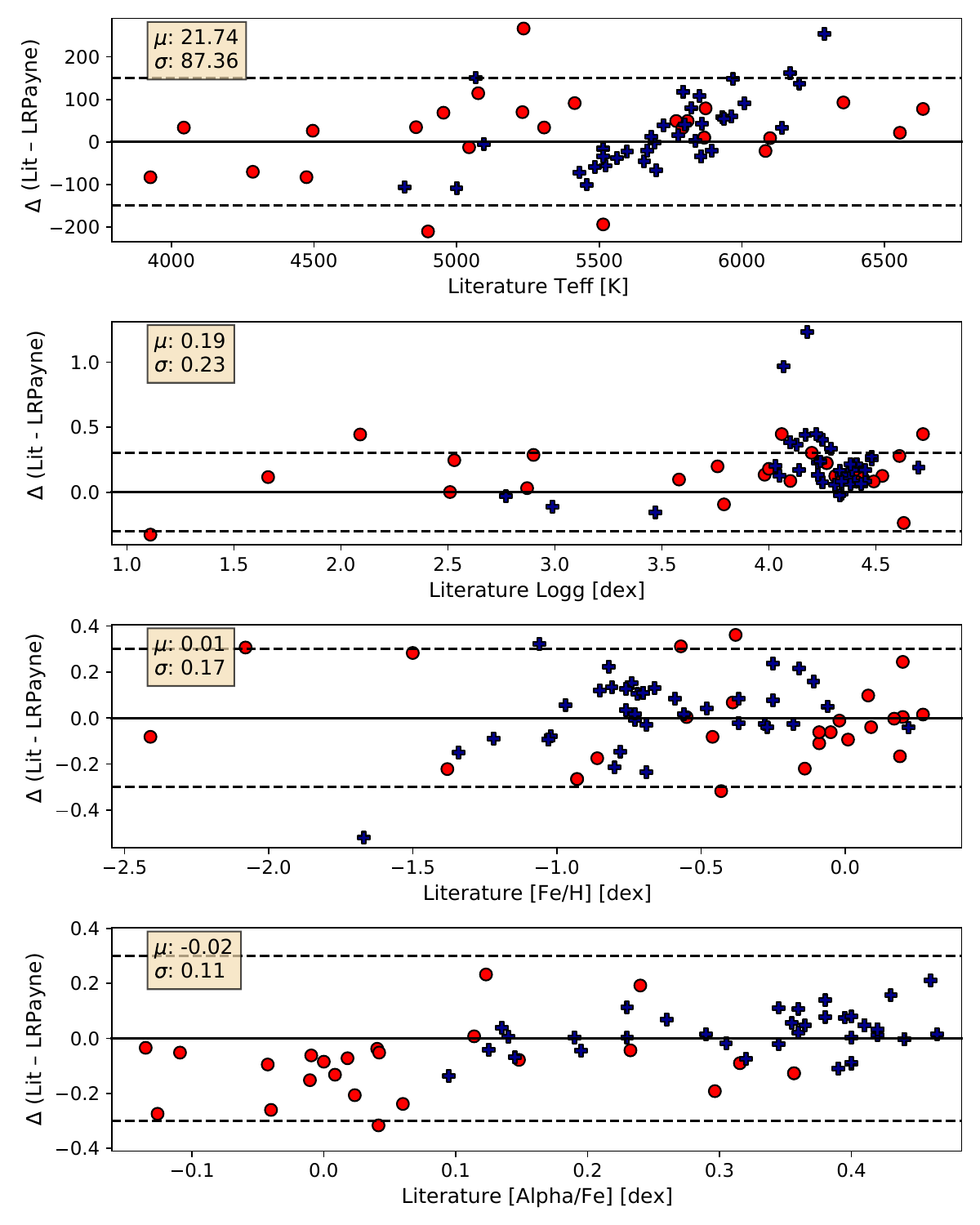}
    \caption{Parameters inferred for Gaia FGK benchmark (red circle) and metal-poor stars (blue plus sign) using LRPayne in comparison with the literature. The solid black line represents zero difference between LRPayne derived value and literature, whereas the two dashed black line represents a range within which the accuracy of LRPayne is comparable to traditional methods of analysis. Literature parameters for benchmark stars were obtained from \cite{jofre2014} and \cite{heiter2015}, and for metal-poor stars from \cite{bensby2014}. The mean and 1$\sigma$ of the distribution is given on the top left of each panel. }
    \label{fig:parameters}
\end{figure}

In Figure \ref{fig:parameters}, we compare the results obtained from LRPayne with those obtained from literature sources, \cite{jofre2014} and \cite{heiter2015} for benchmark stars (red circle), and \cite{bensby2014} for metal-poor stars (blue plus signs). \cite{heiter2015} derived effective temperature using spectral energy distribution, interferometry and distances, which was in turn used alongside luminosity and stellar masses derived through stellar evolutionary models to calculate the surface gravity of the star. \cite{jofre2014} derived the metallicities using spectroscopic analysis of high-resolution spectroscopic data from instruments such as UVES, NARVAL and HARPS (data from the spectral library \citep{blanco2014}). They combined results from six different methods using the same atmospheric models and linelist. \cite{bensby2014} used equivalent widths and excitation equilibrium method for Fe {\sc i} and Fe {\sc ii} lines to estimate the stellar parameters using high-resolution spectroscopy and then apply a NLTE correction on the derived parameters. The main reasons for selecting a study based on NLTE parameters to validate LTE based LRPayne is: 1. one of few studies to homogenously study, with high-resolution spectroscopy, a large sample of metal-poor stars using instruments covering our desired wavelength range of 4200 to 6900 $\AA$; 2. estimates NLTE abundance of Na and Mg, two abundances of interest for Galactic Archeology studies of the outer-halo using WEAVE LR spectra; 3. NLTE corrections applied to most stars are quite small (see Figure 6 in \cite{bensby2014}); 4. Reduced spectra of stars observed with HARPS was readily available at ESO archive; 5. As a bonus, includes abundances such as O, Al, Zn and Y, which are not available for benchmark stars, therefore helps in validation of LRPayne on these elements. Before the comparison, we have to get all the parameters to the same solar scale as \cite{jofre2014} uses a solar iron abundance from \cite{Grevesse2007} [logA(Fe)$_{\odot}$ = 7.45 dex], \cite{bensby2014} uses a solar iron abundance of logA(Fe)$_{\odot}$ = 7.58 dex, and our study is based on \cite{Asplund2009} with [logA(Fe)$_{\odot}$ = 7.50 dex]. Therefore, we add 0.05 dex and subtract 0.08 dex to metallicities of benchmark and metal-poor stars, respectively. 

In Figure \ref{fig:parameters} and \ref{fig:abundances}, taking into account the accuracy and uncertainties associated with parameter estimation at R = 5000 using traditional method of excitation equilibrium of Fe lines, we have defined a range for each stellar label within which an estimate is considered satisfactory. For T$_{\rm eff}$, log(g), [Fe/H], and [X/Fe], this is set to $\pm\, 150\,$K, $\pm\,0.3\,$dex, $\pm\,0.3\,$dex and $\pm\,0.3\,$dex, respectively \citep{2007A&A...467.1373R}. We find a good agreement for most stars with our validation sample. We obtain a difference of 22$\pm$87 K, 0.19$\pm$0.23 dex, 0.01$\pm$0.17 dex and -0.02$\pm$0.11 dex for T$_{\rm eff}$, log(g), [Fe/H] and [$\alpha$/Fe], respectively. Even though there is overall good agreement in T$_{\rm eff}$, we see a clear trend in the values of metal-poor stars. The trend between 5500 K to 6500 K can be attributed to the NLTE vs LTE difference, because, as explained in \cite{bensby2014}, even if the NLTE correction was negligible for stars with T$_{\rm eff}$ < 5500 K, the scatter of correction values increases for hotter stars, which is in line with the trend seen in Figure \ref{fig:parameters}. Among the four parameters, we find a large mean and stand deviation in log(g), but it is driven by the two extreme outlier stars with a $\Delta\,\rm log(g)\,\sim$ 1.0 dex. The two stars HIP22068 [6200K, 4.18dex, -1.34dex] and HIP60632 [6140K, 4.07dex, -1.67dex] are both hot metal-poor dwarfs. We see a similar trend on few other hot metal-poor dwarfs such as HD196892, HD97320 and HD84937, where log(g) is under estimated but to a lesser extent ($\Delta\,\rm log(g)\,\sim$ 0.40 dex). A major reason contributing to the bias is the comparison of LTE results with NLTE values. In Figure 6 of \cite{bensby2014}, the change of log(g) from LTE to NLTE is between 0.1 and 0.3 dex for stars with [Fe/H] $\le$ -1.0 dex. Three of the above mentioned discrepant stars are affected by this, and fall into the satisfactory range when the NLTE effect is taken into account. Along with this, a small contribution could also come from the scarcity of good Fe {\sc ii} lines as they are key for an accurate log(g) determination. At these temperature and metallicity, almost all of the Fe {\sc ii} lines are extremely weak for the neural network to fit properly and therefore usually underestimates the log(g). Even in [$\alpha$/Fe], we see a trend with overestimation for [$\alpha$/Fe]-poor stars and underestimation for [$\alpha$/Fe]-rich stars. However, this is not significant because of the uncertainties associated with literature values ($\sim$ 0.1 dex). Note, our validation sample does not contain any metal-poor giants. Therefore, a modicum of caution is advised when using LRPayne to analyse stars with surface gravity less than 2.5 and metallicity less than -1.3 dex. 

\subsubsection{Abundances}\label{Sec:abun}

\begin{figure*}
    \centering
    \includegraphics[width=0.99\linewidth]{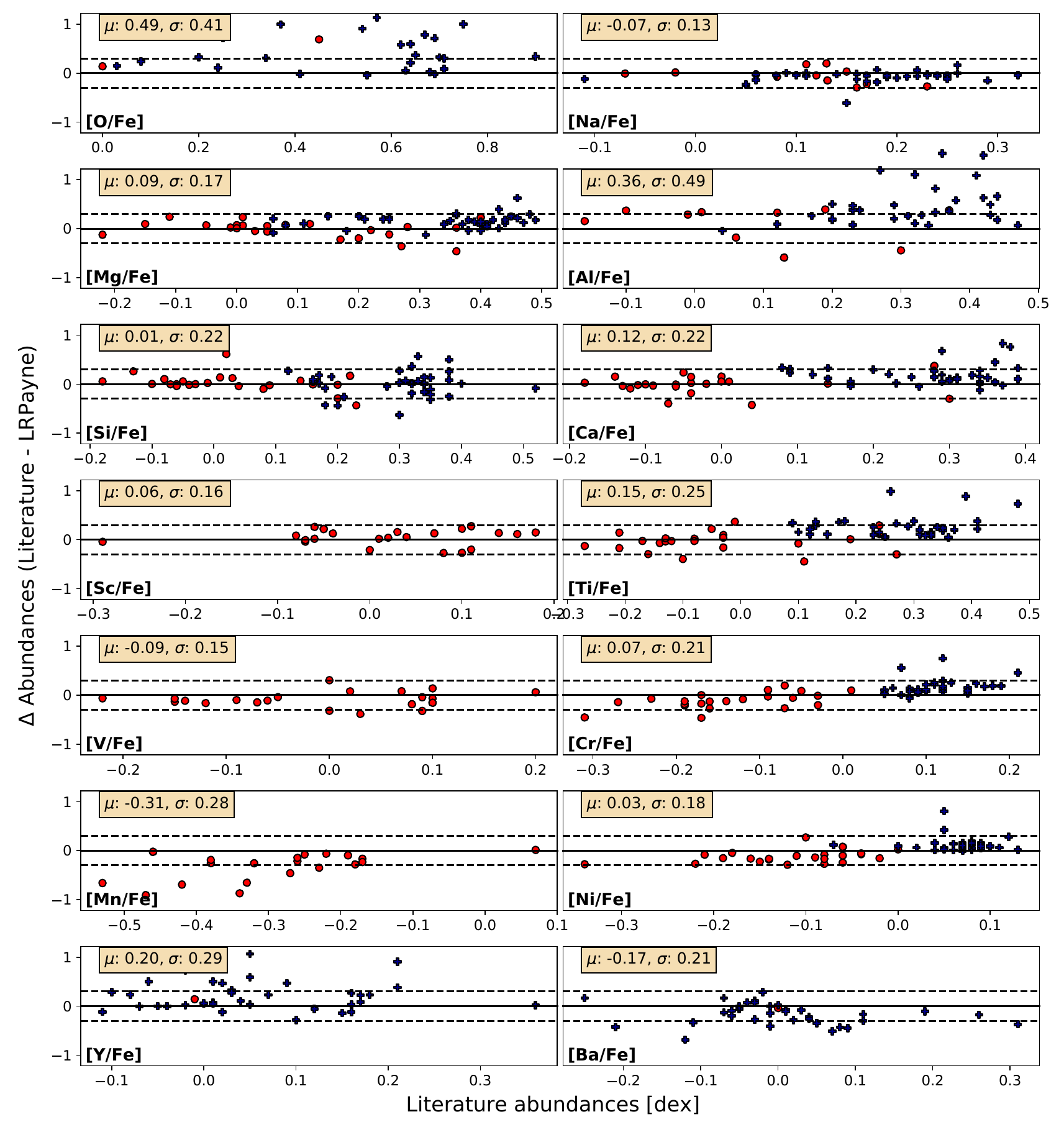}
    \caption{Comparison of abundances for 14 elements obtained from LRPayne and the literature. The red circles represent benchmark stars and dark blue plus signs represent metal-poor stars. Solid black line shows difference between literature and LRPayne equals zero while the dashed black lines represent $\Delta\,=\pm0.3$ dex.}
    \label{fig:abundances}
\end{figure*}

Even though LRPayne has been trained to simultaneously fit 24 elemental abundances, we are able to reliably verify its performance on actual data for only 14 of them. This is because, there is almost no study that performs a homogenous analysis of such a diverse types of stars (in terms of coverage of the parameter space) and also determines abundances for elements such as Li, Na, $\alpha$-, Fe-peak, s-process, and r-process elements. Therefore, we limit our validation to the abundances present in \cite{bensby2014}, \cite{jofre2015} and \cite{jofre2017}. In Figure \ref{fig:abundances}, we compare the abundances obtained from LRPayne with that of literature. 

\subsubsection{$\alpha$ elements}
Among the $\alpha$ elements, we verify five abundances, i.e. O, Mg, Si, Ti and Ca. We like to note, even though Ti is not an $\alpha$ element, it is included in this section due to it behaving similar to an $\alpha$ element. Oxygen is generally a difficult element to analyse within the optical spectrum due to absence of strong CO and OH molecular bands, therefore, when available, most studies use infrared spectrum. Within our desired wavelength range, the two forbidden oxygen lines at 6300 and 6363 $\AA$ are most commonly used to derive the oxygen abundance. But they have two problems: 1. The two lines are in regions affected by telluc absorption, thus depending on the radial velocity of the star, the lines can easily get blended with telluric lines and 2. both lines are generally quite weak with the line at 6363 $\AA$ below the detection limit in most stars (depending on stellar parameters) even at very high SNR. A third difficulty is the NLTE correction (roughly 0.1 dex) applied to these abundances in \cite{bensby2014}. Even with these limitations, LRPayne is still able to recover the oxygen abundance for 18 of the 31 stars or about 60 $\%$ of the sample (after taking into account the uncertainties on the literature values, which is on average of 0.15 dex). When the NLTE correction is taken into account, 20 stars fall within the satisfactory range. Even with majority of the stars having been fitted well, the rest 13 stars are highly discrepant, which leads to a large mean offset of 0.49 dex. Only a portion of the large discrepancy can be explained by a combination of uncertainties on the literature values and NLTE corrections applied by \cite{bensby2014} and the rest is attributed to the difficulties associated with measuring oxygen abundance in optical spectrum. 

Within the other four $\alpha$ elements, LRPayne performs the best on [Si/Fe] with a mean offset of 0.01 dex. We also find LRPayne has a better accuracy on benchmark stars compared to the metal-poor ones. However, when considering the applicable uncertainties, all of the stars except two fall within $\pm$ 0.3 dex of the literature value. For [Mg/Fe], the mean offset is larger than [Si/Fe] at 0.09 dex but with an average uncertainties of 0.09 dex on the literature values, they are in good agreement with the literature. For [Ti/Fe] and [Ca/Fe], the mean offset is slightly larger than [Mg/Fe] at 0.15 and 0.12 dex, respectively. In [Ti/Fe], most stars have good agreement with literature when uncertainties are taken into account, except for three metal-poor stars which are discrepant by more than 0.5 dex and contributes to increase in mean offset. [Ca/Fe] follows the same trend as [Ti/Fe], where most stars are in good agreement with literature except for three. The three stars discrepant in both of the abundances are the same stars, i.e. HIP22068, HIP60631 and HD104006, among which two of them are hot metal-poor dwarf stars. The discrepancy in the estimation of their log(g) will lead to discrepancy in Ti and Ca.

\subsubsection{Odd-Z element} 
Within odd-Z elements, we verify Na and Al using abundances taken from \cite{bensby2014}. All of the sodium abundances from \cite{bensby2014} have been corrected for NLTE effects. But given these corrections were negligible, LRPayne has been able to reliably recover the abundances except for one star. The mean offset ($\mu$) for [Na/Fe] is -0.07 dex, which is close to the average uncertainty of 0.05 dex associated with literature values. Thus, there is a good agreement between the two. However, given that Al has very few suitable lines for abundance analysis that are also weak and blended, it is quite a tricky element for accurate abundance measurement. We find a similar case with LRPayne, as it struggles to accurately fit [Al/Fe] in many stars. Even though about 19 stars or about 48 $\%$ of stars are within $\pm$ 0.3 dex of the literature value, roughly 25 $\%$ of the stars have a discrepancy of larger than 0.5 dex. Upon looking deeper into the discrepant stars, we find a direct correlation between discrepancy with literature values and metallicity of the stars, i.e. more metal-poor a star, higher is the discrepancy. This is expected as a lower metallicities leads to weaker spectral lines and for an element such as Al, the few suitable lines become considerably weak. We also see a trend between the accuracy of LRPayne on [Al/Fe] and log(g) of the star, with higher log(g), i.e. dwarf stars, having poor accuracy.

\subsubsection{Fe-peak elements} 
Within the Fe-peak elements, we investigate five abundances, i.e. Sc, V Cr, Mn and Ni. Among these elements, \cite{bensby2014} only has Cr and Ni for the metal-poor stars, thus for Sc, V and Mn, we only look at the benchmark stars. The overall accuracy of LRPayne on Fe-peak elements is good except for Mn. The other four abundances have excellent agreement with the literature values when the errors are taken into account. Only two stars are found to be discrepant in [Cr/Fe] and [Ni/Fe], which are the same hot metal-poor dwarfs stars having discrepancy in log(g) and the $\alpha$ elements. The mean offset of [Cr/Fe] and [Ni/Fe] is 0.07 and 0.04 dex, respectively. This is smaller than the average uncertainty associated with the literature values. For [V/Fe] and [Sc/Fe], all the stars are within $\pm$ 0.3 dex of the literature value, and with uncertainties, most of the values are easily within 1 $\sigma$ of the literature value. However, for [Mn/Fe], LRPayne seems to recover the abundance reliably only for stars with [Mn/Fe] > -0.25 dex. The reliability reduces significantly after [Mn/Fe] < -0.25 dex, with only three out of nine stars being within $\pm$ 0.3 dex of the literature value. We also find the stars that have high discrepancy are all metal-rich giants and stars within $\pm$ 0.3 dex of literature values are all dwarf stars. The most possible reason for such a behaviour of LRPayne is the saturation of the Mn lines, which is more likely in giant stars than dwarfs. Such saturation effects remove any patterns between the elemental abundance and the spectral line strength, thus making it difficult for the neural network to reproduce the underlying correlation.

\subsubsection{Heavy elements}
Among the heavy elements, we are able to verify two s-process elements Ba and Y using the metal-poor stars. Even though, from Figure \ref{fig:abundances} it seems like there an offset (in case of [Ba/Fe]) and a large scatter (in case of [Y/Fe]), more than  
70 $\%$ of the stars fall within $\pm$ 0.3 dex of the literature value upon consideration of the uncertainties on the literature values. This is because of the large average uncertainties of 0.10 and 0.16 dex associated with the literature values of [Ba/Fe] and [Y/Fe].

\subsubsection{Uncertainties}

Within the workflow of LRPayne, there are two sources of uncertainties associated to the parameter estimation. One is the fitting error of the $\chi^{2}$ minimisation process while another stems from the variation in the trained ANN models due to random initialization during each training run.

\begin{figure*}
\sidecaption
    \centering
    \includegraphics[width=12cm]{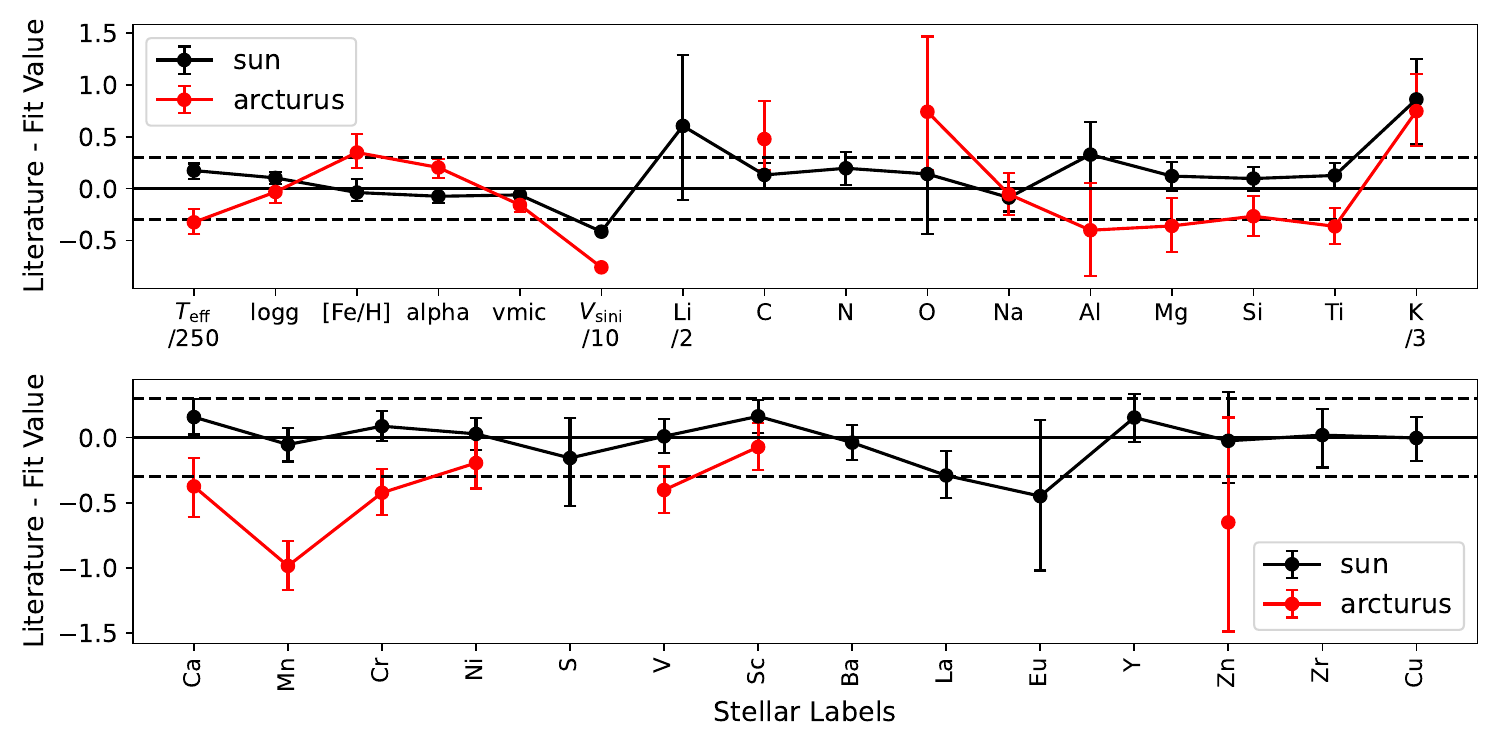}
    \caption{Uncertainties on stellar labels for Sun (black) and Arcturus (sun). Missing values for Arcturus is due to unavailability of literature values. The uncertainties shown are from LRPayne with the literature errors assumed to be zero to visually represent the fitting errors. }
    \label{fig:mcmcerror}
\end{figure*}

In order to probe both of these sources, we employed a mcmc-like methodology. We train 10 different ANN models using different shuffling of the training set but keeping all the hyper-parameters the same. For each of the 10 models, a star goes through the entire workflow 2000 times with the initial guess for the $\chi^{2}$ minimisation process being the only change. This is repeated for all the 10 models, with a total of 20000 fitting procedure performed per star. This provides a posterior distribution for each of the parameters which can then be used to measure the uncertainties with 1$\sigma$ corresponding to 16th and 86th percentile of the distribution. This method takes both the sources of uncertainties into consideration: fitting error is considered by repeated the $\chi^{2}$ minimisation 2000 times and variations in trained models is considered by fitting the same star with 10 different models. The uncertainties have been estimated for 10 stars that cover a wide region of the parameter space. The limited number is due to large computation time required for the uncertainty analysis of each star. Figure \ref{fig:mcmcerror} shows the uncertainties associated with the stellar labels in Sun (S/N $\sim$ 500) and Arcturus (S/N $\sim$ 500). For most stellar labels, the uncertainties are reasonable with errors on stellar parameters being quite good for both the stars. The largest uncertainties are obtained for elements such as Li, O, Al, K, S, Eu and Zn which are generally difficult to estimate in low-resolution spectrum. Apart from these elements, the average uncertainties on abundances is about 0.13 dex for Sun and 0.21 dex for Arcturus. An interesting trend about the uncertainties is that, in all the stellar parameters and abundances that are fitted well by LRPayne, the errors associated with Arcturus are larger than the errors associated with the Sun. This could be a result of giant stars having a lot more spectral features in their spectra compared to a dwarf stars and thus leading to an increase in interpolation error. In Table \ref{tab:error}, we summarise the uncertainties on different stellar labels depending on the type of star. 

\begin{table*}[]
\centering
\caption{Uncertainties on stellar labels on stars with different stellar parameters. Errors marked in bold represent large uncertainties ($\sigma$ > 0.5 dex), and represent elements that cannot be measured precisely. }
\label{tab:error}
\resizebox{\textwidth}{!}{%
\begin{tabular}{cccccccccccccccc}
\hline
\textbf{Star} & \textbf{eT$_{\rm eff}$} & \textbf{elog(g)} & \textbf{e[Fe/H]} & \textbf{e[$\alpha$/Fe]} & \textbf{eV$_{\rm mic}$} & \textbf{eV$_{\rm sini}$} & \textbf{e[Li/Fe]} & \textbf{e[C/Fe]} & \textbf{e[N/Fe]} & \textbf{e[O/Fe]} & \textbf{e[Na/Fe]} & \textbf{e[Al/Fe]} & \textbf{e[Mg/Fe]} & \textbf{e[Si/Fe]} & \textbf{e[Ti/Fe]} \\
\hline
\hline
\textbf{Sun}     & 19 & 0.07& 0.07& 0.07& 0.06& 0.41         & \textbf{1.38}& 0.06& 0.11& \textbf{0.60} & 0.09& 0.29         & 0.09& 0.04& 0.07\\
\textbf{Arcturus}& 32 & 0.10& 0.17& 0.09& 0.07& 0.27         & \textbf{0.95}& 0.32& 0.43 & \textbf{0.64} & 0.12& 0.43& 0.20& 0.10& 0.06\\
\textbf{muLeo}   & 30 & 0.07& 0.15& 0.04& 0.11& 0.01         & \textbf{0.50}& 0.24& 0.18& 0.48 & 0.07& 0.14         & 0.15& 0.09& 0.08\\
\textbf{HD101259}& 51 & 0.13& 0.17& 0.07& 0.12& 3.59         & \textbf{1.45}& 0.20& 0.19& \textbf{0.91} & 0.08& 0.34& 0.17& 0.05& 0.07\\
\textbf{HIP38134}& 46 & 0.19& 0.30& 0.11& 0.27& 3.32         & \textbf{1.31}& 0.12& \textbf{0.69} & \textbf{1.21} & 0.09& \textbf{0.90}& 0.43 & 0.16& 0.13\\
\textbf{HD107328}& 42 & 0.14& 0.20& 0.11& 0.09& 0.20         & \textbf{1.26}& 0.20& 0.29& 0.48 & 0.14& \textbf{0.68}& 0.28& 0.11& 0.06\\
\textbf{HD104006}& 50 & 0.16& 0.31& 0.13& 0.26& 2.62         & \textbf{1.30}& 0.23& \textbf{0.55} & 0.46 & 0.15& 0.49& 0.30& 0.25& 0.11\\
\textbf{HD126681}& 50 & 0.08& 0.13& 0.06& 0.17& 1.39         & \textbf{1.28}& 0.09& 0.31 & \textbf{0.79} & 0.13& \textbf{0.51}& 0.18& 0.12& 0.07\\
\textbf{HD140283}& 49 & 0.16& 0.21& 0.13& 0.11& 3.88         & \textbf{1.67}& 0.21& \textbf{0.78} & \textbf{1.04} & 0.14& \textbf{0.56}& 0.19& 0.22& 0.07\\ \hline
\textbf{Star} & \textbf{e[K/Fe]} & \textbf{e[Ca/Fe]} & \textbf{e[Mn/Fe]} & \textbf{e[Cr/Fe]} & \textbf{e[Ni/Fe]} & \textbf{e[S/Fe]} & \textbf{e[V/Fe]} & \textbf{e[Sc/Fe]} & \textbf{e[Ba/Fe]} & \textbf{e[La/Fe]} & \textbf{e[Eu/Fe]} & \textbf{e[Y/Fe]} & \textbf{e[Zn/Fe]} & \textbf{e[Zr/Fe]} & \textbf{e[Cu/Fe]} \\
\hline
\textbf{Sun}     & \textbf{1.27}& 0.08& 0.06& 0.03& 0.06& 0.32& 0.07         & 0.06& 0.08& 0.15& \textbf{0.57} & 0.14         & 0.34 & 0.20& 0.13\\
\textbf{Arcturus}& \textbf{1.03}& 0.15& 0.10& 0.07& 0.11& \textbf{1.08}& 0.07         & 0.08& 0.14& 0.17& 0.17& 0.19         & \textbf{0.84} & 0.07& 0.24\\
\textbf{muLeo}   & \textbf{0.89}& 0.15& 0.14& 0.07& 0.09& 0.24& 0.07         & 0.09& 0.19& 0.15& 0.42 & 0.16         & \textbf{0.73} & 0.20& 0.39 \\
\textbf{HD101259}& \textbf{1.33}& 0.13& 0.08& 0.08& 0.11& \textbf{0.83}& 0.13         & 0.11& 0.19& 0.01& \textbf{0.51} & 0.25         & \textbf{1.24} & 0.29& 0.46 \\
\textbf{HIP38134}& \textbf{1.42}& 0.43 & 0.14& 0.13& 0.11& \textbf{0.86}& 0.26         & 0.13& 0.24& 0.00& \textbf{1.01} & 0.28         & \textbf{1.27} & \textbf{0.51} & \textbf{0.93} \\
\textbf{HD107328}& \textbf{1.51}& 0.18& 0.11& 0.07& 0.14& \textbf{1.13}& 0.09         & 0.12& 0.21& 0.25& 0.21& 0.28         & \textbf{1.20} & 0.14& 0.42 \\
\textbf{HD104006}& \textbf{1.25}& 0.37 & 0.17& 0.11& 0.16& \textbf{1.49}& 0.17         & 0.19& 0.18& 0.00& \textbf{0.80} & 0.36& \textbf{1.34} & \textbf{0.73} & 0.40 \\
\textbf{HD126681}& \textbf{1.19}& 0.19& 0.11& 0.07& 0.05& \textbf{0.62}& 0.24         & 0.16& 0.25& 0.44 & \textbf{0.73} & 0.26         & \textbf{1.15} & \textbf{0.57} & \textbf{0.52} \\
\textbf{HD140283}& \textbf{1.34}& 0.18& 0.06& 0.08& 0.09& \textbf{1.36}& 0.13         & 0.13& 0.01& 0.17& \textbf{0.81} & 0.07         & \textbf{1.04} & 0.41 & 0.31 \\
\hline
\end{tabular}%
}

\vspace{1em}

\resizebox{\textwidth}{!}{%
\begin{tabular}{llll}
\textbf{Sun:} 5770\,K, 4.40\,dex, 0.00\,dex &
\textbf{Arcturus:} 4286\,K, 1.66\,dex, -0.52\,dex &
\textbf{muLeo:} 4474\,K, 2.51\,dex, 0.25\,dex &
\textbf{HD101259:} 5000\,K, 2.99\,dex, -0.82\,dex \\
\textbf{HIP38134:} 5837\,K, 4.23\,dex, -0.89\,dex &
\textbf{HD107328:} 4496\,K, 2.09\,dex, -0.33\,dex &
\textbf{HD104006:} 5067\,K, 4.71\,dex, -0.77\,dex &
\textbf{HD126681:} 5484\,K, 4.48\,dex, -1.30\,dex \\
\textbf{HD140283:} 5514\,K, 3.58\,dex, -2.36\,dex & & &
\end{tabular}%
}
\end{table*}

\subsubsection{Dependency on S/N}
\begin{figure*}
\sidecaption
    \centering
    \includegraphics[width=12cm]{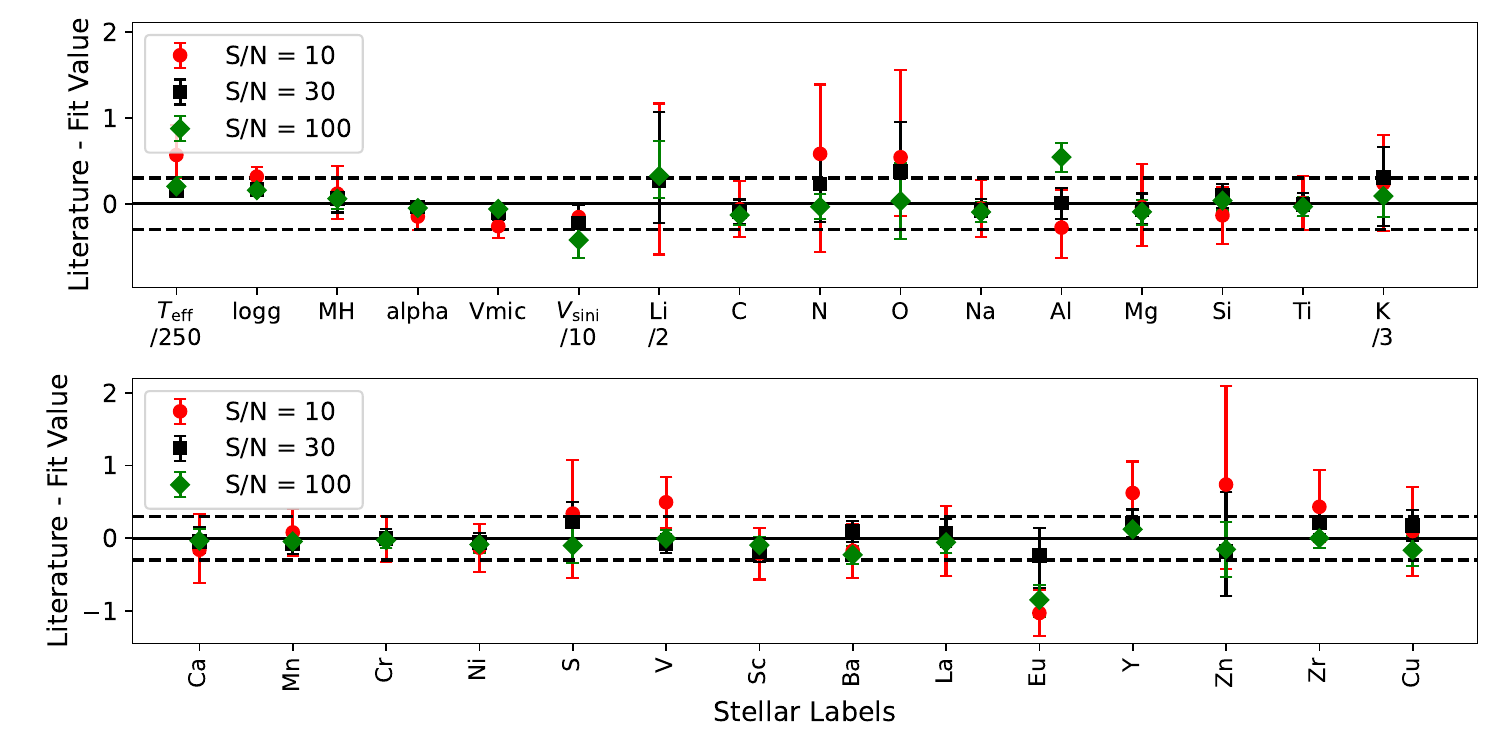}
    \caption{Comparison of fitted values from LRPayne with the literature values for Sun spectra at three different S/N, S/N = 10 (red circles), S/N = 30 (black square) and S/N = 100 (green diamond). The uncertainties shown are from LRPayne with the literature errors assumed to be zero to visually represent the fitting errors.  }
    \label{fig:snr}
\end{figure*}

As our primary objective is to use LRPayne to analyse low-resolution data of a spectroscopic survey mission, it is important to understand how the quality of the spectrum affects the reliability of LRPayne. Generally, as surveys try to find a balance between number of observed stars and quality of those observations, the data very rarely tends to have a high S/N (> 100 per $\AA$). To test this, we use LRPayne to fit three solar spectra with different amounts of noise added to them. The S/N of the three spectra is 10, 30 and 100. The results of the fitting is shown in Figure \ref{fig:snr}. From the figure, its evident that the reliability of LRPayne upon fitting a 30 S/N spectrum is almost the same as fitting a 100 SNR spectrum. Only at very low S/N of 10, we see the results deviating more from the literature values. This result shows that LRPayne is highly capable of handling spectral data with lower S/N. 

\section{Summary}
In this paper, we present a neural network-based algorithm specifically designed for the analysis of low-resolution optical spectra from large-scale galactic surveys called LRPayne. Our algorithm is an adaptation of {\em The Payne} and employs a fully connected artificial neural network with three hidden layers. The training set comprised of 70,000 synthetic stellar spectra generated using Turbospectrum code within iSpec using 1D MARCS atmosphere models and a modified (GES + VALD) linelist. The algorithm analyses spectra degraded to $R = 5000$ with a wavelength range of 4200 - 6900 $\AA$, making it ideal for surveys such as WEAVE. Our key findings are summarized below:

\begin{itemize}
\item Technical Performance: Internal accuracy tests demonstrate that LRPayne achieves excellent interpolation accuracy, with median errors of less than 0.13 $\%$ for 90 $\%$ of the synthetic validation sample. The algorithm successfully recovers input stellar parameters and abundances from synthetic spectra, even at signal-to-noise ratios as low as 20, though with expected degradation in accuracy for challenging elements such as Li, K, and N.

\item Stellar Parameters: Validation on 64 real stars (23 Gaia FGK benchmark stars and 41 metal-poor stars) reveals robust performance for fundamental stellar parameters. We achieve mean differences of $22 \pm 87$~K for effective temperature, $0.19 \pm 0.23$~dex for surface gravity, and $0.01 \pm 0.17$~dex for metallicity when compared to literature values. The algorithm shows particular strength in determining effective temperatures and metallicities across a wide range of stellar types.

\item Chemical Abundances: LRPayne demonstrates reliable abundance determination for multiple elements across different nucleosynthetic groups. Elements such as Na, Mg, Si, and most Fe-peak elements (Cr, Ni, V, Sc) have typical accuracies of 0.1--0.2~dex. The algorithm successfully determines abundances for $\alpha$-elements (Mg, Si, Ti, Ca), Fe-peak elements (Sc, V, Cr, Ni), and heavy elements (Ba, Y), providing crucial information for Galactic archaeology studies.

\item Challenges and Limitations: We identify specific challenges in the analysis of certain stellar types and elements. Oxygen abundance determination remains difficult due to the weakness of optical oxygen lines and telluric contamination. Manganese abundances show systematic biases in metal-rich giants, likely due to line saturation effects. Aluminium abundances are challenging in metal-poor stars and dwarfs due to weak spectral features. Surface gravity determination for hot metal-poor dwarfs shows systematic underestimation, partly attributable to LTE versus NLTE effects and the weakness of Fe{\sc ii} lines in this parameter regime. Another important challenge is the substantial adverse effect of synthetic gap on the performance of the neural network. Such a gap can easily derail the analysis when not taken into consideration, leading to incorrect results and conclusions. Masking the wavelength pixels is a easy way to correct for this, but for future, there might be a need to use a more sophisticated method of dealing with the synthetic gap (to ensure minimal information loss). Lastly, even though the performance of LRPayne on the verification sample has been good, it is still a small sample. A more comprehensive analysis is needed to fully understand the behaviour of LRPayne, but this will only be possible with commencement of surveys such as WEAVE and 4MOST.   

\item Signal-to-Noise Performance: LRPayne maintains robust performance down to S/N $\sim$ 30, with only modest degradation compared to high S/N spectra. This capability is crucial for survey applications where achieving high S/N for all targets is impractical.

\item Immediate Application: The algorithm demonstrates particular efficacy in determining abundances of key elements such as Na and Mg, which are crucial tracers for identifying first-generation (1G) and second-generation (2G) stellar populations within Galactic globular clusters. We plan use LRPayne to analyse WEAVE low-resolution data to identify and characterise these populations to understand their formation and evolution within the clusters. 

\end{itemize}

\addcontentsline{toc}{chapter}{Bibliography}
\bibliographystyle{aa}
\bibliography{reference.bib}

\end{document}